
\input harvmac

\def\Titleh#1#2{\nopagenumbers\abstractfont\hsize=\hstitle\rightline{#1}%
\vskip .5in\centerline{\titlefont #2}\abstractfont\vskip .5in\pageno=0}

\def\CTPa{\it Center for Theoretical Physics, Department of Physics,
      Texas A\&M University}
\def\CTPb{\it College Station, TX 77843-4242, USA}
\def\HARCa{\it Astroparticle Physics Group,
Houston Advanced Research Center (HARC)}
\def\HARCb{\it The Woodlands, TX 77381, USA}

\def\ie{\hbox{\it i.e.}}     
\def\eg{\hbox{\it e.g.}}

\catcode`\@=11 

\def\lsim{\mathrel{\mathpalette\@versim<}}
\def\gsim{\mathrel{\mathpalette\@versim>}}
\def\@versim#1#2{\vcenter{\offinterlineskip
    \ialign{$\m@th#1\hfil##\hfil$\crcr#2\crcr\sim\crcr } }}
\def\boxit#1{\vbox{\hrule\hbox{\vrule\kern3pt
      \vbox{\kern3pt#1\kern3pt}\kern3pt\vrule}\hrule}}

\def\etal{{\it et. al.}}

\def\t1{{\tilde 1}}
\def\ov{\overline}

\def\JL{J. L. Lopez}
\def\DVN{D. V. Nanopoulos}

\def\GeV{\,{\rm GeV}}
\def\TeV{\,{\rm TeV}}

\def\NPB#1#2#3{Nucl. Phys. B {\bf#1} (19#2) #3}
\def\PLB#1#2#3{Phys. Lett. B {\bf#1} (19#2) #3}

\def\PRD#1#2#3{Phys. Rev. D {\bf#1} (19#2) #3}
\def\PRL#1#2#3{Phys. Rev. Lett. {\bf#1} (19#2) #3}
\def\PRT#1#2#3{Phys. Rep. {\bf#1} (19#2) #3}

\def\TAMU#1{Texas A \& M University preprint CTP-TAMU-#1}

\nref\EKN{J. Ellis, S. Kelley and D. V.  Nanopoulos, \PLB{249}{90}{441},
\PLB{260}{91}{131}, \NPB{373}{92}{55}; P. Langacker and M.-X. Luo,
\PRD{44}{91}{817}; F. Anselmo, L. Cifarelli, A. Peterman, and A. Zichichi,
Nuovo Cim. {\bf104A} (1991) 1817 and {\bf105A} (1992) 581.}
\nref\EN{J. Ellis and \DVN, \PLB{110}{82}{44}.}
\nref\HKT{For a recent reappraisal see \eg, J. Hagelin, S. Kelley, and
T. Tanaka, MIU-THP-92/59.}
\nref\bsg{S. Bertolini, F. Borzumati, A. Masiero, and G. Ridolfi,
\NPB{353}{91}{591}; V. Barger, M. Berger, and R. J. N. Phillips,
\PRL{70}{93}{1368}; J. Hewett, \PRL{70}{93}{1045}; R. Barbieri and G. Giudice,
CERN-TH.6830/93; \JL, \DVN, and G.T. Park, \TAMU{16/93}.}
\nref\Dickreview{For reviews see: R. Arnowitt and P. Nath, {\it Applied N=1
Supergravity} (World Scientific, Singapore 1983);
H. P. Nilles, \PRT{110}{84}{1}.}
\nref\LNZb{\JL, \DVN, and A. Zichichi, CERN-P-TH.6667/92, CTP-TAMU-68/92.}
\nref\aspects{S. Kelley, \JL, \DVN, H. Pois, and K. Yuan, \TAMU{16/92}
and CERN-TH.6498/92 (to appear in Nucl. Phys. B).}
\nref\LNZa{\JL, \DVN, and A. Zichichi, \PLB{291}{92}{255}.}
\nref\LNP{\JL, \DVN, and H. Pois, \PRD{47}{93}{2468}.}
\nref\LNPZ{\JL, \DVN, H. Pois, and A. Zichichi, \PLB{299}{93}{262}.}
\nref\LNWZ{\JL, \DVN, X. Wang, and A. Zichichi, CERN/PPE/92-194,
CTP-TAMU-76/92.}
\nref\LNPWZ{\JL, \DVN, H. Pois, X. Wang, and A. Zichichi, CERN-PPE/93-16,
CTP-TAMU-89/92.}
\nref\hera{\JL, \DVN, X. Wang, and A. Zichichi, CERN-PPE/93-64,
CTP-TAMU-15/93.}
\nref\oldEW{E. Eliasson, \PLB{147}{84}{67}; S. Lim, \etal, \PRD{29}{84}{1488};
J. Grifols and J. Sola, \NPB{253}{85}{47}; B. Lynn, \etal, in Physics at LEP,
eds. J. Ellis and R. Peccei, CERN Yellow Report CERN86-02, Vol. 1;
R. Barbieri et al., \NPB{341}{90}{309}.}
\nref\Bilal{A. Bilal, J. Ellis, and G. Fogli, \PLB{246}{90}{459}.}
\nref\DHY{M. Drees, K. Hagiwara, and A. Yamada, \PRD{45}{92}{1725}.}
\nref\DH{M. Drees and K. Hagiwara, \PRD{42}{90}{1709}.}
\nref\BFC{R. Barbieri, M. Frigeni, and F. Caravaglios, \PLB{279}{92}{169}.}
\nref\ABC{G. Altarelli, R. Barbieri, and F. Caravaglios, CERN-TH.6770/93.}
\nref\AN{M. Matsumoto, J. Arafune, H. Tanaka, and K. Shiraishi,
\PRD{46}{92}{3966}; R. Arnowitt and P. Nath, \PRL{69}{92}{725}; P. Nath and
R. Arnowitt, \PLB{287}{92}{89} and \PLB{289}{92}{368}.}
\nref\LN{For a review see, A. B. Lahanas and \DVN,
\PRT{145}{87}{1}.}
\nref\LNPWZh{\JL, \DVN, H. Pois, X. Wang, and A. Zichichi, CERN-PPE/93-17,
CTP-TAMU-05/93 (to appear in Phys. Lett. B).}
\nref\PDG{Particle Data Group, \PRD{45}{92}{S1}.}
\nref\Kennedy{D. Kennedy and B. Lynn, \NPB{322}{89}{1};
D. Kennedy, B. Lynn, C. Im, and R. Stuart, \NPB{321}{89}{83}.}
\nref\PT{M. Peskin and T. Takeuchi, \PRL{65}{90}{964};
W. Marciano and J. Rosner, \PRL{65}{90}{2963};
D. Kennedy and P. Langacker, \PRL{65}{90}{2967}.}
\nref\efflagr{B. Holdom and J. Terning, \PLB{247}{90}{88};
M. Golden and L. Randall, \NPB{361}{91}{3};
A. Dobado, D. Espriu, and M. Herrero, \PLB{255}{91}{405}.}
\nref\AB{G. Altarelli and R. Barbieri, \PLB{253}{90}{161}.}
\nref\ABJ{G. Altarelli, R. Barbieri, and S. Jadach, \NPB{369}{92}{3}.}
\nref\BB{R. Barbieri, CERN Report No. CERN-TH.6659/92 (unpublished).}
\nref\MSBAR{S. Sarantakos, A. Sirlin, and W. J. Marciano, \NPB{217}{83}{84}.}
\nref\ONSHELL{A. Sirlin, \PRD{22}{80}{971}.}
\nref\LMKN{C. S. Lim, T. Inami, and N. Sakai, \PRD{29}{84}{1488};
B. A. Kniehl, \NPB{352}{91}{1}.}
\nref\Lynn{D. Levinthal, F. Bird, R. Stuart, and B. Lynn, Z. Phys. C, {\bf 53}
(1992) 617.}
\nref\VernonHARC{See, for example, V. Barger, talk presented at the HARC
workshop ``Recent
Advances in the Superworld", The Woodlands, Texas, April 1993.}

\nfig\I{The total contribution to $\epsilon_1$ as a function
of the lightest chargino mass $m_{\chi^\pm_1}$ (upper row)
and also as a function of the lightest Higgs boson mass $m_h$ (lower row)
for the minimal $SU(5)$ supergravity model. Points between the
two horizontal solid lines are allowed at $90\%$ CL.}
\nfig\II{Same as in Fig. 1 except for the no-scale flipped $SU(5)$ supergravity
model. The three distinct curves (from lowest to highest) correspond to
\hfuzz=10pt $m_t=100,130,160 \GeV$.}
\nfig\III{The total contribution to $\epsilon_2$ as a function
of the lightest chargino mass $m_{\chi^\pm_1}$ (upper row)
and also as a function of the lightest Higgs boson mass $m_h$ (lower row)
for the minimal $SU(5)$ supergravity model. Points between the
two horizontal solid lines are allowed at $90\%$ CL.}
\nfig\IV{Same as in Fig. 3 except for the no-scale flipped $SU(5)$ supergravity
model.}
\nfig\V{The total contribution to $\epsilon_3$ as a function
of the lightest chargino mass $m_{\chi^\pm_1}$ (upper row)
and also as a function of the lightest Higgs boson mass $m_h$ (lower row)
for the minimal $SU(5)$ supergravity model. Points between the
two horizontal solid lines are allowed at $90\%$ CL.}
\nfig\VI{Same as in Fig. 5 except for the no-scale flipped $SU(5)$ supergravity
model.}

\Titleh{\vbox{\baselineskip12pt\hbox{CTP--TAMU--19/93}
\hbox{ACT--07/93}}}
{\vbox{\centerline{Precision Electroweak Tests of the Minimal}
        \vskip2pt\centerline{and Flipped SU(5) Supergravity Models}}}
\centerline{JORGE L. LOPEZ, D.~V.~NANOPOULOS, GYE T. PARK, H. POIS, and KAJIA
YUAN}
\bigskip
\centerline{\CTPa}
\centerline{\CTPb}
\centerline{and}
\centerline{\HARCa}
\centerline{\HARCb}
\vskip .5in
\centerline{ABSTRACT}
We explore the one-loop electroweak radiative corrections in the minimal
$SU(5)$ and the no-scale flipped $SU(5)$ supergravity models via explicit
calculation of vacuum polarization contributions to the $\epsilon_{1,2,3}$
parameters. Experimentally, $\epsilon_{1,2,3}$ are obtained from a global fit
to the LEP observables, and $M_W/M_Z$ measurements. We include $q^2$-dependent
effects which have been neglected in most previous ``model-independent"
analyses of this type. These effects induce a large systematic negative shift
on $\epsilon_{1,2,3}$ for light chargino masses ($m_{\chi^\pm_1}\lsim70\GeV$).
In agreement with previous general arguments, we find that for increasingly
large sparticle masses, the heavy sector of both models rapidly decouples, \ie,
the values for $\epsilon_{1,2,3}$ quickly asymptote to the Standard Model
values with a {\it light} Higgs ($m_{H_{SM}}\sim100\GeV$). Specifically, at
present the $90\%$ CL upper limit on the top-quark mass is $m_t\lsim175\GeV$
in the no-scale flipped $SU(5)$ supergravity model. These bounds can be
strengthened for increasing chargino masses in the $50-100\GeV$ interval.
In particular, for $m_t\gsim160\GeV$, the Tevatron may be able to probe
 through gluino($\tilde g$) and squark($\tilde q$) production up to
$m_{\tilde g}\approx m_{\tilde q}\approx250\GeV$, exploring at least half
of the parameter space in this model.

\bigskip
\Date{April, 1993}

\newsec{Introduction}

Despite the lack of a single {\it direct} piece of experimental evidence for
any supersymmetric partner to the Standard Model (SM) particles, there are some
remarkable {\it indirect} results which indicate that superpartners may be
operative at the ${\cal O}(1\TeV)$ scale, and involve several predictions and
constraints which could be regarded as much more than coincidence. The most
dramatic one is the high-precision unification of the running gauge couplings
at superhigh energies in the minimal $SU(5)$ supergravity model, only when
virtual sparticle effects intervene at the $\TeV$ scale and modify the running
of the gauge couplings \EKN. Virtual supersymmetric effects have long been
known to be a possible menace to supersymmetric models because of the
potentially large one-loop induced flavor-changing-neutral-current (FCNC)
processes such as $K-\bar K$ mixing and CP violation in the $K$ system \EN,
requiring high degeneracy of squark masses. Supergravity models with universal
soft-supersymmetry breaking naturally tame these effects \HKT. More recently,
one-loop supersymmetric contributions to FCNC process $b\to s\gamma$ have
been shown to be a possible deep probe of supersymmetric models \bsg.

In this work we explore yet another avenue for experimentally testable virtual
effects in supergravity models, namely one-loop electroweak (EW) radiative
corrections. In particular, we explore the minimal $SU(5)$ \Dickreview\ and the
no-scale flipped $SU(5)$ supergravity models \LNZb\ which can be considered to
be  prototype traditional versus string-inspired supergravity models. This work
represents a continuation of a general program which we have initiated in the
study of supergravity models. Previously, we have explored the broader issues
of EW radiative symmetry breaking, and bounds on the parameter space of
supergravity models resulting from the many experimental and consistency
constraints \aspects. More recently, we have included the stringent constraints
from proton decay and the cosmological neutralino relic density, applicable to
the minimal $SU(5)$ model we consider here \refs{\LNZa,\LNP}. Finally, we have
explored chargino-neutralino production and detection at Fermilab \LNWZ,
chargino, neutralino, slepton, and Higgs production and detection at LEPII
\LNPWZ, and selectron-neutralino and sneutrino-chargino production and
detection at HERA \hera. Here, we present a complete study of one-loop EW
radiative corrections in the minimal and flipped $SU(5)$ supergravity models,
incorporating the most recent global fits to the precision measurements from
LEP.

Several previous studies of EW radiative corrections in the generic minimal
supersymmetric standard model (MSSM) and supergravity models exist
\refs{\oldEW,\Bilal,\DHY,\DH,\BFC,\ABC}. In many cases however, the emphases
were different and the contributions of $q^2$-dependent effects have often been
neglected under the assumption that they are unimportant, as would be the case
for $m_{susy}>M_Z$. In the MSSM however, it has been recently demonstrated that
for very light charginos
($m_{\chi^\pm}\lsim 60-70\GeV$), a significant $Z$-boson wave-function
renormalization threshold effect can modify the results dramatically \BFC.
In the models we consider here, this effect leads to strong correlations
between the chargino and the top-quark mass. Specifically, we find that at
present the $90\%$ CL upper limit on the top-quark mass is $m_t\lsim175\GeV$ in
the no-scale flipped $SU(5)$ supergravity model. These bounds
can be strengthened for increasing chargino masses in the $50-100\GeV$
interval. For example, in the flipped model for
$m_{\chi^\pm_1}\gsim60\,(70)\GeV$,
we find $m_t\lsim165\,(160)\GeV$. As expected, the heavy sector of both models
decouples quite rapidly with increasing sparticle masses, and at present, only
$\epsilon_1$ leads to constraints on the parameter spaces of these models.
However, as we discuss, future values for $\epsilon_2,\epsilon_3$ may also be
constraining. Finally, an upper limit to $m_t$ leads to upper limits to the
lightest CP-even Higgs boson mass ($m_h$) which may be within reach of LEPII.

In the following section, we briefly discuss the overall motivation and
structure of the minimal and flipped $SU(5)$ models that we study here.
In Sec. 3, we present an overview of EW radiative corrections, and justify our
use of the $\epsilon_{1,2,3}$ parameterization. In Sec. 4 we discuss the
specific minimal and flipped $SU(5)$ supergravity model contributions to
the vacuum polarization diagrams, and the resultant contributions to the
$\epsilon_{1,2,3}$, and finally conclude in Section 5.

\newsec{The minimal and flipped SU(5) supergravity models}

The minimal \Dickreview\ and no-scale flipped \LNZb\ $SU(5)$ supergravity
models that we consider here can be regarded as prototypes for realistic
traditional versus string-inspired supergravity unified models. At low energy,
both contain (i) the SM $SU(3)\times SU(2)\times U(1)$ gauge symmetry which is
radiatively broken at the weak scale, and (ii) the three SM generations and two
Higgs doublets of matter representations at the EW scale (along with their
superpartners). There are however, several crucial differences between the two
models: (i) each unifies into a larger (but different) gauge group, namely
$SU(5)$ versus $SU(5)\times U(1)$, (ii) the
unification scale in the minimal $SU(5)$ model is $M_U\approx10^{16}\GeV$,
whereas in the flipped case, the (string) unification scale is
$M_U\approx10^{18}\GeV$, close to the Planck scale. Unification at this higher
scale is due to the effects of additional, vector-like representations
($Q,\ov Q,D,\ov D$) that naturally fit into the ${\bf 10,\ov {10}}$ flipped
$SU(5)$ representations. At the unification scale the heavy field content is
also quite different, and leads to different conclusions regarding proton
decay: in the minimal $SU(5)$ model proton decay is highly constraining
\refs{\AN,\LNP,\LNPZ}, whereas in the flipped model it is not. Finally, (iii)
the pattern of soft-supersymmetry breaking at the unification scale is quite
different: in the flipped case, the sole source of SUSY breaking is due to a
universal gaugino mass ($m_{1/2}$) as is typical in unified no-scale
supergravity models \LN, whereas in the minimal $SU(5)$ model, universal scalar
($m_0$) and trilinear ($A$) contributions must also be included.

The constraints of gauge and Yukawa unification along with the SUSY breaking
assumptions and the satisfaction of all the consistency and phenomenological
constraints on these models lead to a restricted five-dimensional parameter
space in this class of models \aspects. Besides the three SUSY breaking
parameters ($m_{1/2},\xi_0\equiv m_0/m_{1/2},\xi_A\equiv A/m_{1/2}$;
$\xi_0=\xi_A=0$ in the flipped model) one also has the top-quark mass ($m_t$),
and the ratio of Higgs vacuum expectation values ($\tan\beta=v_2/v_1$) defined
at the EW scale. The sign of the superpotential Higgs mixing term $\mu$ is also
undetermined, and both cases ($\pm|\mu|$) must be considered. Primarily as a
result of the quite different pattern of SUSY breaking, the low-energy
predictions for the squark, slepton, chargino and neutralino mass spectra are
quite distinct, and lead to strikingly different phenomenology in the two
models (see Refs. \refs{\LNZa,\LNP,\LNPZ,\LNWZ,\LNPWZ,\hera,\LNPWZh} for
detailed discussions).

As described in detail in Ref. \aspects, our strategy for studying these
models involves a discrete sampling of the
($m_{1/2},\xi_0,\xi_A,\tan\beta,m_t$) parameter space over their allowed
domain. Several consistency and phenomenological constraints restrict
the range of the model parameters and yield an allowed region in parameter
space. See Refs. \LNPZ\ and \LNZb\ for the determination of the parameter
spaces in the minimal and flipped $SU(5)$ supergravity models respectively.
In the remainder of this work we explore the constraints arising from
one-loop EW radiative corrections.

\newsec{One-loop EW radiative corrections and the $\epsilon_{1,2,3}$
parameters}
It is now well known that quantum effects in the EW sector lead to significant,
measurable corrections to the various tree-level EW parameters which are
otherwise in discrepancy with the data by more than $2\sigma$. For example,
predictions for $\sin^2\theta_w$ and $M_W$ which are obtained from  various
$Z$-scale measurements and low-energy neutrino scattering experiments are
consistent only if
one-loop effects are considered. In the SM (assuming a single Higgs doublet to
effect the EW symmetry breaking), there is a residual $SU_V(2)$ (custodial)
symmetry which protects the tree-level relation $\rho\equiv
M^2_Z\cos^2\theta_w/M^2_W=1$. This symmetry is broken at the loop level by
unequal top- and bottom-quark Yukawa couplings and hypercharge interactions
(or equivalently $\sin^2\theta_W\not=0$) leading to the well known quadratic
$m_t$ and logarithmic $m_{H_{SM}}$ dependences of
$\delta\rho$.\foot{Supersymmetry breaking effects introduce new sources of
explicit $SU(2)_V$ breaking, and so does $\tan\beta\not=1$. See Ref. \DH\ for
a detailed discussion of this point.} These effects typically do not decouple
with increasingly larger mass scales. Global SM fits to all of the low-energy
and electroweak data therefore constrain $m_t<194,178,165\GeV$ for
$m_{H_{SM}}=1000,250,50\GeV$ at the $90\%$ CL respectively \PDG.

Recently, several schemes have been introduced which effectively parametrize
the EW vacuum polarization corrections \refs{\Kennedy,\PT,\efflagr,\AB}. One
can easily show that an expansion of the vacuum polarization tensors to order
$q^2$, results in three independent physical parameters. Alternatively, one can
show from an effective field theory point of view, that there are three
additional terms in the lagrangian \efflagr. In the $(S,T,U)$ scheme \PT, a SM
reference value for $m_t,m_{H_{SM}}$ is used, and the deviation from this
reference is calculated and is considered to be ``new" physics. This scheme is
only valid to lowest order in $q^2$, and is therefore not applicable to a
theory with new, light $(\sim M_Z)$ particles. In the two supergravity
models we explore here, each point in parameter space is actually a distinct
model, and a SM reference point is not meaningful. For these reasons, we choose
to use the scheme of Refs. \refs{\AB,\ABJ,\BFC} where the contributions are
{\it absolute} and valid to higher order in $q^2$. This $\epsilon_{1,2,3}$
scheme is therefore more applicable to a global fit of the supergravity models
we consider here. Regardless of the scheme used, all of the global fits to the
three physical parameters are {\it entirely consistent} with the SM at $90\%$
CL.

In principle, every observable, such as the $Z$ widths into a pair of fermions
($\Gamma_f,\ f=l,b$), the forward-backward asymmetries at the $Z$-pole
($A^f_{FB},\ f=l,b$), the $\tau$-polarization asymmetry ($A^\tau_{Pol}$), the
ratio of neutral- to charged-current processes in deep-inelastic neutrino
scattering on nuclei ($R_\nu$), etc., is a distinct measurable and is not
directly related to the others. Various assumptions need to be made in order to
combine these measurements in a way which is predictive and sensitive to new
physics effects. The first assumption usually made is quark-lepton
universality. Secondly, the dominant ``new" contributions are assumed to arise
from the process-independent (\ie, ``oblique") vacuum polarization amplitudes.
In fact, even if there are non-negligible non-oblique (\ie, vertex and box
diagrams) corrections, the set of observables can be restricted such that these
contributions are minimized. Such is the case of the vertex corrections to the
$Z$-$b$-$\bar b$ coupling which are important in supersymmetric models but
their impact is minimized by considering $A^b_{FB}$ where these mainly cancel
out \BB. It follows that the $\Gamma_b$ observable should not be included if
the vacuum-polarization dominance assumption is to hold.\foot{Low-energy
measurements such as $R_\nu$
and the weak charge $Q_W$ measured in atomic parity violation in cesium are
not included since non-oblique supersymmetric corrections down to low-energies
may be quite important in the presence of a light chargino.} The above
assumptions allow for a large set of measurables to be combined into a global
fit.

In the $\epsilon_{1,2,3}$ scheme these parameters\foot{In our calculations, we
use $\ov{MS}$ scheme \MSBAR\ throughout, where one has an advantage of having
some expressions simpler than in the on-shell scheme \ONSHELL.} can be written
as follows \refs{\BFC,\BB},
\eqna\I
$$\eqalignno{\epsilon_1&=e_1-e_5-{\delta G_{V,B}\over G}-4\delta g_A,&\I a\cr
\epsilon_2&=e_2-s^2 e_4-c^2 e_5-{\delta G_{V,B}\over G}
				-\delta g_V-3 \delta g_A,&\I b\cr
\epsilon_3&=e_3+c^2 e_4-c^2 e_5+{c^2-s^2\over 2 s^2}\delta g_V
         -{1+2s^2\over 2 s^2}\delta g_A,&\I c\cr}$$
where $s^2=1-c^2=\sin^2\theta_W$ and the $e_i$, ($i=1\to5$) are the following
combinations of the vacuum polarization amplitudes
\eqna\II
$$\eqalignno{
e_1&={\alpha\over 4\pi s^2 M^2_W}[\Pi^{33}_T(0)-\Pi^{11}_T(0)],&\II a\cr
e_2&=F_{WW}(M_W^2)-{\alpha\over 4\pi s^2}F_{33}(M_Z^2),&\II b\cr
e_3&={\alpha\over 4\pi s^2}[F_{3Q}(M_Z^2)-F_{33}(M_Z^2)],&\II c\cr
e_4&=F_{\gamma\gamma}(0)-F_{\gamma\gamma}(M_Z^2),&\II d\cr
e_5&= M_Z^2F^\prime_{ZZ}(M_Z^2),&\II e\cr}$$
and the $q^2\not=0$ contributions $F_{ij}(q^2)$ are defined by
\eqn\III{\Pi^{ij}_T(q^2)=\Pi^{ij}_T(0)+q^2F_{ij}(q^2).}
The $\delta g_{V,A}$ in Eqns. \I{a-c} are the contributions to the vector and
axial-vector form factors at $q^2=M^2_Z$ in the $Z\to l^+l^-$ vertex from
proper vertex diagrams and fermion self-energies, and $\delta G_{V,B}$ in Eqns.
\I{a,b} comes from the one-loop box, vertex and fermion self-energy corrections
to the $\mu$-decay amplitude at zero external momentum. It is important to note
that these non-oblique SM corrections are non-negligible, and must be included
in order to obtain an accurate SM prediction. Our method consists of making a
graphical fit to the SM curves in Ref. \BFC. In the following section we
calculate the vacuum polarization contributions to $\epsilon_{1,2,3}$ in the
two models we consider. As discussed above, we assume throughout that the
non-oblique supersymmetric contributions to the measurables that are included
in the global fit are negligible.

\newsec{Contributions from the minimal and flipped SU(5) supergravity models}
It is well known in the MSSM that the largest contributions to $\epsilon_1$
(\ie, $\delta\rho$ if $q^2$-dependent effects are neglected) are expected to
arise from the $\tilde t$-$\tilde b$ sector, and in the limiting case of a very
light stop, the contribution is comparable to that of the $t$-$b$ sector \DH.
The remaining squark, slepton, chargino, neutralino, and Higgs sectors all
typically contribute considerably less. For increasing sparticle masses, the
heavy sector of the theory decouples, and only SM effects  with a {\it light}
Higgs survive. However, for very light chargino, we have mentioned that a
$Z$-wavefunction renormalization threshold effect can introduce a substantial
$q^2$-dependence
in the calculation, thus modifying significantly the standard $\delta\rho$
results.  For completeness, we include the
complete vacuum polarization contributions from the Higgs sector, the
supersymmetric chargino-neutralino and sfermion sectors, and also the
corresponding contributions in the SM. Our analytical expressions for the
$\Pi^{ij}_T(q^2)$ agree with those given in Refs. \refs{\DHY,\DH} as well as
other existing references \LMKN. For the SUSY vacuum polarization
contributions, new infinities are introduced which must cancel in the full
calculation of any physical, gauge invariant observable.
Divergence-cancellation is thus a crucial and useful check of the masses and
couplings, especially in the chargino-neutralino sector where the cancellation
is not obvious. We have verified this consistency condition both analytically
and numerically. Following the convention of Ref. \BFC, we include the running
of the electric charge from $q^2=0$ up to $q^2=M^2_Z$, due to light quarks and
leptons, by using $\alpha(M_Z)$ in Eq. \II{}. This implies that one should use
$s^2=0.2312$ \refs{\BFC,\BB}. Moreover, $e_4$ should include only
contributions from the remaining charged particles, \ie, from $W, t$, and the
supersymmetric charged particles.

In Fig. 1 (2) we show the calculated values of $\epsilon_1$ versus the lightest
chargino mass ($m_{\chi^\pm_1}$) and versus the lightest CP-even Higgs boson
mass ($m_h$), for the sampled points in the minimal (flipped) $SU(5)$
supergravity model. In Fig. 2 (the flipped case) three representative values of
$m_t$ were used, $m_t=100,130,160\GeV$, whereas in Fig. 1 (the minimal $SU(5)$
case), several other values for $m_t$ in the range $90\GeV\le m_t\le 160\GeV$
were
sampled. In both models, but most clearly in the flipped model
(Fig. 2) one can see how quickly the sparticle spectrum decouples as
$m_{\chi^\pm_1}$ increases, and the value of $\epsilon_1$ asymptotes to the SM
value appropriate to each value of $m_t$ and for a {\it light} ($\sim 100
\GeV$) Higgs mass. This result is in agreement with Ref. \LNPWZh\ where
we have shown that the Higgs sector is SM-like for virtually all points in
these models. The threshold effect of $\chi^\pm_1$ is manifest as
$m_{\chi^\pm_1}\rightarrow {1\over2} M_Z$ and is especially visible for $\mu<0$
in both models. This systematic, negative contribution comes from the
$F^\prime$ term in $e_5$ which is considered to be a $Z$-wave-function
renormalization effect \BFC. Note from Eq. (3.1) that all
$\epsilon$'s are affected similarly.
Due to the presence of the $\chi^+-\chi^-$ cut
near $M_Z$ the $F^\prime$ term is not expected to be very accurate as
$m_{\chi^\pm_1}\rightarrow {1\over 2} M_Z$. However, according to Ref. \BFC,
for $m_{\chi^\pm_1}>50\GeV$, this correction agrees to better than $10\%$
with the one obtained in a more accurate way. Our numerical results for
$\epsilon_1$ can be compared with Ref. \BFC\ where the authors calculated
$\epsilon_1$ in the MSSM neglecting squark contributions for different choices
of $m_{\chi^\pm_1}$ and assumed a common slepton mass $m_S$. The maximum
wavefunction renormalization effect in Ref. \BFC\ occurs at
$m_{\chi^\pm_1}=m_S=50\GeV$. Our results for the two realistic supergravity
models at fixed $m_{t}$ and $m_{\chi^\pm_1}$ fall between their minimum and
maximum values.

Recent values for $\epsilon_{1,2,3}$ obtained from a global fit to the
LEP (\ie, $\Gamma_l,A^{l,b}_{FB},A^\tau_{pol})$ and $M_W/M_Z$ measurements
are \BB,
\eqn\IV{\epsilon_1=(-0.9\pm 3.7)10^{-3},\quad\epsilon_2=(9.9\pm
8.0)10^{-3},\quad\epsilon_3=(-0.9\pm 4.1)10^{-3}.}
For $\epsilon_1$ it is clear that virtually all the sampled points in the
minimal $SU(5)$ supergravity model (Fig. 1) are within the $\pm 1.64 \sigma$
($90\%$ CL) bounds (denoted by the two horizontal solid lines in the figures).
Since several values for $90\le m_t\le 160\GeV$ were sampled, the trends for
fixed $m_t$ are not very clear from the figure. Nonetheless, the points
just outside the $1.64 \sigma$ line correspond to $m_t=160\GeV$, which
are therefore excluded at the $90\%$ CL. We note that we have imposed the
improved experimental constraint $m_h>60\GeV$ \LNPWZh, which has the effect of
removing many of the points corresponding to very light $\chi^\pm_1$,
particularly for $\mu>0$. We have also imposed $m_{\chi^\pm_1}\gsim 50\GeV$ as
an accuracy cut due to the threshold effect mentioned above. Larger values for
$m_t$ were not explored since they are not expected to be consistent
with the combined proton decay and cosmological constraints in this model \LNP.
Thus, no useful upper bound on $m_t$ (from $\epsilon_1$) can be obtained in
this model. However, should these two constraints be relaxed, we would expect
to obtain upper bounds similar to those that follow for the flipped model
below.

In the flipped model (Fig. 2), the upper bound on $m_t$ depends sensitively
on the chargino mass. For example, for $m_t=160\GeV$, only light chargino
masses would be acceptable at $90\%$ CL. In fact,  we have scanned the region
$130\GeV\le m_t\le 190\GeV$ in increments of $5\GeV$ and obtained the maximum
values
for $m_{\chi^\pm_1}$ allowed by the experimental value for $\epsilon_1$ at
$90\%$ CL. These are given in the Table I. One can immediately see the strong
correlation between $m_t$ and $m_{\chi^\pm_1}$: as $m_t$ rises, the upper limit
to $m_{\chi^\pm_1}$ falls, and vice versa. In particular, for $m_t\le150\GeV$
all values of $m_{\chi^\pm_1}$ are allowed, while one could have $m_t$ as large
as $160\,(175)\GeV$ for $\mu>0\,(\mu<0)$ if the chargino mass were light
enough.

Turning to the other two variables, in Figs. 3-6, one can see that both
$\epsilon_2$ and $\epsilon_3$ do not constrain the models at the present level
of experimental accuracy. Nonetheless, it is evident from the figures that both
$\epsilon_2,\epsilon_3$ are also affected by the threshold effect for light
$\chi^\pm_1$, since $\epsilon_{2,3}$ depend on $e_5$ in a similar way that
$\epsilon_1$ does (see Eq. \I{}). The practice of plotting
$\epsilon_1,\epsilon_3$ and drawing the $90\%$ CL correlated error ellipse
(see \eg, Refs. \refs{\ABC,\BB}) may prove useful when $\epsilon_3$ becomes
constraining, and at present it is not.

In Figs. 3-6, the bottom row presents $\epsilon_{2,3}$ versus $m_h$.  Since
one-loop corrections to $m_h$ depend strongly on $m_t$, there is an upper limit
to $m_h$ for fixed $m_t$. This upper limit is most apparent in the flipped case
(Fig. 2) for the three choices of $m_t$. For $\mu>0$, the $90\%$ upper limit to
$\epsilon_1$ (see Fig. 2, bottom row) requires $m_h\lsim 100 \GeV$
(for $m_t=160 \GeV$), whereas for
$\mu<0$ the limit is $m_h\lsim 105 \GeV$. The {\it absolute} upper limit
to $m_h$ would correspond to a smaller top mass
($\simeq 150 \GeV$) and would be close to $m_h\sim 115 \GeV$.
In the minimal $SU(5)$ model (see
Fig. 1, bottom row), there are at present no useful limits to $m_h$ from
$\epsilon_1$ (\eg, for $\mu<0$, $m_h\lsim94\GeV$ as opposed to
$m_h\lsim97\GeV$ if the $\epsilon_1$ constraint is not imposed). For each
$m_t$ branch in Fig. 2 (bottom row) one can see the drop in $\epsilon_1$
vs. $m_h$ corresponding to points where $m_{\chi^\pm_1}$ is light.
Since there is no $Zhh$ coupling (due to bose symmetry), there is no
threshold effect for light $h$ Higgs masses ($m_h\simeq 60 \GeV$). The
apparent drop in $\epsilon_1$ vs. $m_h$ ($\mu<0$) in Fig. 1 is due to the
fact that $m_h$ and $m_{\chi_1^\pm}$ are somewhat correlated (since both
scale by $m_{1/2}$). The relationship between $\epsilon_{2,3}$ vs. $m_h$ is
qualitatively similar to $\epsilon_1$ in the flipped case,
except for a closer spacing between the three sets of $m_t$ values due to
a milder $m_t$ dependence (see bottom row in Figs. 4,6).
As is
well known, $\epsilon_{2,3}$ depend on the top-quark mass only as
$\ln(m_t/M_Z)$, and are therefore more sensitive to new physics. This can be
seen most clearly in Figs. 4,6 for $\epsilon_{2,3}$ for the flipped case, where
the curves corresponding to the three $m_t$ values are grouped very close
together. However, all are well within the $90\%$ CL experimental limits. This
weak dependence on $m_t$ makes these variables potentially good probes of new
physics, especially if the top-quark is not within the reach of the Tevatron.
For the future, it is expected that $2\times 10^6$ Z-events/experiment at LEPI
will lower the uncertainties for $\epsilon_{1,3}$ to $\pm 2.0\times 10^{-3}$
\BB. Depending on the new central values, one may be able to make definitive
statements about constraints on the parameter spaces of these models.

One should be careful when making statements about upper limits to $m_t$ based
on the oft-quoted global fits to the data. There is an inherent uncertainty in
the fits due to the fact that the correlation matrix between the various
measurables is almost always omitted from the analysis (see however Ref.
\Lynn). Nonetheless, the general trends we find here can be expected to hold:
for very light chargino mass ($m_{\chi^\pm_1}$), the top quark can be heavy
($m_t\gsim 160 \GeV$) and thus possibly escape detection at Fermilab from the
data
collected during 1993-94 (assuming ${\cal L}=75\;{\rm pb}^{-1}$ of integrated
luminosity). However, light charginos should be readily detectable at LEPII
in both models \LNPWZ, at the Tevatron in the minimal $SU(5)$ model \LNWZ, and
at HERA in the flipped $SU(5)$ model \hera. Moreover, since $m_t\gsim160\GeV$
requires $m_{\chi^\pm_1}\lsim70\GeV$ in the flipped model (see Table I)
, upper bounds on many other sparticle masses follow, since all scale with
$m_{1/2}$.
These bounds are shown in Table II. Note that with $100\;{\rm pb}^{-1}$
of integrated luminosity, the Tevatron may be able to probe up to
$m_{\tilde g}\approx m_{\tilde q}\approx250\GeV$ through the missing $p_T+$
jets signal \VernonHARC. Thus, the full gluino and squark mass range (for the
$\mu>0$) may be accessible at the Tevatron if the top quark is not seen.

\newsec{Conclusions}

In this work, we have explored the minimal $SU(5)$  and the no-scale flipped
$SU(5)$ supergravity model contributions to the three physical one-loop EW
correction parameters, and have compared these results to the latest global fit
to the LEP precision measurements. We include the {\it complete} sparticle
spectrum, and also include $q^2$-dependent effects which are important for a
light chargino spectrum. This effect is due to a Z-boson wavefunction
renormalization threshold effect, and systematically lowers the values of
$\epsilon_{1,2,3}$. As we have shown, the effect is most significant for
$m_{\chi^\pm_1}\lsim 70\GeV$, where light $\chi^\pm_1$ can reduce the value for
$\epsilon_1$ to fall within the experimental limits for $m_t$ as large as
$175\GeV$ in the flipped model. However, this region for $\chi^\pm_1$ is quite
restrictive, and should  be thoroughly explored at LEP II \LNPWZ. Moreover, in
this case the Tevatron should be able to explore at least half of the parameter
space of the flipped model through gluino and squark production.

Conversely, if $m_{\chi^\pm_1}\gsim 70\GeV$ and the top is not seen with ${\cal
L}=100\;{\rm pb}^{-1}$ integrated luminosity at CDF (by early 1995), then the
two models here would be disfavored (at 90\% CL). Only $\epsilon_1$ is
constraining, due to an $m^2_t$ dependence, and the inferred upper limits to
$m_t$ correspond to SM limits with a light $(\sim 100 \GeV)$ Higgs, since both
models decouple quite rapidly with increasing $m_{\chi^\pm_1}$. For the future,
LEP measurements may be able to reduce the errors on the $\epsilon_i$
considerably (e.g. $\epsilon_3$ by $75\%$), and we may begin to constrain the
models. However, if $m_{\chi^\pm_1}\gsim70\GeV$, the constraints on the two
models would apply equally well to the SM: if the SM is eventually ruled out by
precision EW physics, then so are the two supergravity models we consider here.
Considering the Higgs sector, an upper limit to $m_t$ leads to a $90\%$ CL
upper limit $m_h\lsim 115\GeV$ in the flipped model, however the present
errors do not allow for any useful constraints on $m_h$ in the minimal model.

The future of precision EW tests of the minimal and flipped $SU(5)$ models
looks quite promising.  If the SM is ruled out by precision EW tests,
the only ``loophole" for the two supergravity models considered here would be
to include a light chargino. If the tests remain consistent with the SM, then
we can assume that the SUSY spectrum  has decoupled, and would resemble the SM
with a light Higgs, placing a more restrictive upper limit on $m_t$. In this
case, the top quark should be seen at Fermilab in the near future. Thus, in
combination with direct sparticle and top-quark searches, we may be able to
unambiguously test
these models in present and future experiments at Fermilab and LEPII.

\bigskip
\bigskip
\bigskip
\noindent{\it Acknowledgments}: This work has been supported in part by DOE
grant DE-FG05-91-ER-40633. The work of J.L. has been supported by an SSC
Fellowship. The work of D.V.N. has been supported in part by a grant from
Conoco Inc. The work of G.P. and K.Y. has been supported by World-Laboratory
Fellowships. G.P thanks the Phenomenology Institute of the University of
Wisconsin
for his use of its computing facilities for this work.
H.P. wishes to thank B. Lynn and J. White for several helpful
discussions.

\vfill\eject
\listrefs

\input tables
\vbox{\tenpoint\noindent {\bf Table I}: Maximum allowed  chargino mass
($m_{\chi^\pm_1}$) for different $m_{t}$ (in GeV) at $90\% $CL in the flipped
$S
U(5)$ model.
In the entries Y(N) means all points are within (outside) the LEP bounds
at $90\% $CL}
\bigskip
\thicksize=1.0pt
\begintable
$m_{t}$|$\mu>0$|$\mu<0$\cr
145| Y | Y   \nr
150| Y | Y   \nr
155|68| 95 \nr
160|66| 72 \nr
165| N | 63 \nr
170| N | 58 \nr
175| N | 53 \nr
180| N | N   \endtable
\bigskip
\bigskip

\vbox{\tenpoint\noindent {\bf Table II}: Maximum allowed  particle masses (in
GeV) at $90\% $CL for $m_{t}\gsim 160\GeV$  in the flipped $SU(5)$ model.}
\bigskip
\begintable
|$\mu>0$|$\mu<0$\cr
$\chi^0_1$ |$ 35$|$ 38$\nr
$\chi^\pm_1$ |$ 66 $|$ 72$\nr
$\tilde g$|$240$|$330$\nr
$\tilde u_L, \tilde d_L$  |$235$|$322$\nr
$\tilde u_R$  |$216$|$297$\nr
$\tilde d_R$  |$213$|$293$\nr
$\tilde t_1$  |$180$|$195$\nr
$\tilde t_2$  |$320$|$410$\nr
$h$           |$ 100$|$ 105$\nr
$A$           |$ 190$|$ 310$\nr
$H$           |$205$|$ 320$\nr
$H^\pm$         |$205$|$ 325$
                                                                \endtable

\listfigs
\bye